\documentstyle[amstex,amssymb,aps,prl,psfig,newlfont]{revtex}

\begin{document}

\author{Beno\^\i t Gr\'emaud and Dominique Delande}
\address{
Laboratoire Kastler-Brossel, Universit\'e Pierre et Marie Curie,
4 Place Jussieu, 75005 Paris, France}
\title{Photo-ionization of the helium atom close to the double ionization
threshold: towards the Ericson regime}
\date{\today}
\maketitle

\begin{abstract}
We calculate the photo-ionization cross-section from the
ground state of the helium atom, using the complex rotation
method and diagonalization of sparse matrices.
This produces directly the positions and widths of the
doubly excited ${}^1\mathbf{P}^o$ resonances together
with the photo-ionization cross-section.
Our calculations up to the $N=9$ threshold 
are in perfect agreement with recent experimental
data and show the transition from a regular structure at low energy
to a chaotic one at high energy, where various resonances strongly
overlap.
\end{abstract}



The Helium atom is one of the prototype of atomic systems whose
classical dynamics is mainly chaotic and during the past thirty years, 
it has been 
matter of numerous studies from both theoretical~\cite{Fano63,Herrick75,Lin84,%
Feagin88,Ho91,Tang93,Richter92,Wintgen93} and 
experimental~\cite{Madden63,Domke95,Domke96} points of view. 
But, unlike other atomic systems like 
for instance the hydrogen atom in
magnetic field, the effects of chaos are not very well understood and more
profound studies are needed. This requires the resolution
of the full quantum problem. Especially, one has to take into account
all the degrees of freedom and all correlations between the two electrons,
as well as the auto-ionizing character of the doubly excited states.

In this 
letter, we present numerical calculations of the cross-section
of the one-photon photoionization from the ground state of the helium atom
and compare them with the recently obtained high-resolution spectra 
of ${}^1\mathbf{P}^o$ doubly excited states.
The agreement
with the most recent experimental data from references~\cite{Domke95,Domke96} --
up to the $N=9$ ionization threshold, less than 1 eV from the
double ionization threshold and corresponding to 64 open channels --
is excellent for the whole energy range, proving the high efficiency of the
method. Predictions for better experimental resolutions are also given.
We also show that at low energy the resonances can be classified with
respect to the Herrick's $(N,K,T)$ approximate quantum numbers~\cite{Herrick75},
($(N,K)_n$ Lin's simplified notation~\cite{Lin84} will be used hereafter).
At high energy,
this classification progressively breaks down. Eventually, above
the $N=7$ threshold, the various resonances strongly overlap:
the mean energy spacing between consecutive resonances becomes
smaller than their typical width. Oscillations in the photo-ionization 
cross-section
can then no longer be associated with individual resonances: random-like
fluctuations -- known as Ericson fluctuations -- should be observed
in the cross-section.

The quantum Hamiltonian in atomic units 
(\mbox{$\hbar=m_{e^-}=4\pi\epsilon_0=e^2=1$}) is given by:
\begin{equation}
H=\frac {\mathbf{P}^2_1+\mathbf{P}^2_2}2-\frac 2{r_1}-\frac 2{r_2}+
\frac 1{r_{12}}
\end{equation}
\noindent where $\mathbf{P}_i=-\imath\hbar\nabla_i$ is the momentum operator
of electron $i$, $r_i$ its distance to the nucleus and $r_{12}$ the 
inter-electronic distance. All spin-orbits, relativistic and QED effects 
(at most of the order of a fraction of meV) are neglected,
which is consistent with the experimental resolution (of the order of 1 meV).
The corrections due the finite mass of the nucleus are taken into account
by using the effective values for the
double-ionization threshold $I_{\infty}$ and the Rydberg constant $R_{H\!e}$ 
given in the reference~\cite{Domke95}, namely $I_{\infty}=79.003$ eV and 
$R_{H\!e}=13.6038$ eV.

Using the rotational invariance of the Hamiltonian, the angular dependency of
a wavefunction can be factorized as follows~\cite{Brink}:
\begin{equation}
\Psi_{L\,M}=\sum_{T=-L}^L \mathcal{D}^{L^{\displaystyle *}}_{MT}
(\psi,\theta,\phi) \Phi_{T}^{(LM)}(x,y,z)
\end{equation}
\noindent where $(\psi,\theta,\phi)$ are Euler angles defining the 
transformation from the laboratory frame to a molecular-like frame whose
$z'$ axis is the inter-electronic axis. $|T|$ is then the $\Lambda$
($\Sigma$, $\Pi$...)
quantum number in a molecule. The $\mathcal{D}^{L^{\displaystyle *}}_{MT}$
are the wavefunctions of the rigid rotor  and reduce to the usual
spherical harmonics for $T=0$. Finally $(x,y,z)$ are the
perimetric coordinates, symmetric combinations of $r_1$, $r_2$ and $r_{12}$:
\begin{equation}
\left\{
\begin{array}{lr}
x= &r_1+r_2-r_{12} \\
y= &r_1-r_2+r_{12} \\
z=&-r_1+r_2+r_{12}
\end{array} \right.
\end{equation}
For each pair of good quantum numbers $(L,M)$, we obtain an effective
Hamiltonian acting on the different $\Phi_T$'s (coupled by Coriolis-like
terms). 
The two remaining discrete symmetries -- parity and exchange between the two electrons --
are exactly taken
into account by adding constraints on the $\Phi_T$'s\cite{Wintgen93}.

As stated before, above the first ionization threshold, all
states become resonances because of the coupling with the continua 
(autoionizing states). Using the
complex rotation method~\cite{Ho91,Buchl94}, 
we obtain these resonances as complex eigenvalues of a complex
Hamiltonian $H(\theta)$, which is obtained by the replacements 
$\mathbf{r}_i \rightarrow \mathbf{r}_i e^{\imath\theta}$ and 
$\mathbf{P}_i \rightarrow \mathbf{P}_i e^{-\imath\theta}$, where $\theta$ is
a real parameter. 
 The fundamental properties of the spectrum of $H(\theta)$ are the
following~: the continua of $H$ are rotated by an angle $2\theta$ in the
lower complex half-plane around their branching point. Each other complex
eigenvalue $E_i=\epsilon_i - i \Gamma_i/2$ lies in the lower half-plane
and coincides with a resonance of $H$ with energy $\epsilon_i$ and width $\Gamma_i$. 
These quantities are independent of $\theta$ provided that the
complex eigenvalue has been uncovered by the 
rotated continua. The bound states, which 
are resonances with zero width, stay on the real axis. This method also allows
to compute quantities of physical interest, like photo-ionization 
cross-section, probability densities or expectation values of operators (e.g.
$\cos\theta_{12}$), enlighting the contribution of a given
resonance to them. For instance, the cross-section is given by~\cite{Buchl94}~:
\begin{equation}
\label{crosscomp}
\sigma(\omega)=\frac {4\pi\omega}{c}\mathrm{Im}
\sum_i\frac{\langle\overline{E_{i\theta}}|R(\theta)T|g\rangle^2}{E_{i\theta}
-E_g-\hbar\omega}
\end{equation}
\noindent $T$ is the dipole operator, $\hbar\omega$ is the photon energy,
$|g\rangle$ is the ground state (of energy $E_g$). 
$\langle\overline{E_{i\theta}}|$ is the  transpose 
of the eigenvector $|E_{i\theta}\rangle$ of $H(\theta)$ for the  
eigenvalue $E_{i\theta}$ (i.e. the complex conjugate of $\langle E_{i\theta}|$).
$R(\theta)$ is the rotation operator, essential to obtain the right (complex) 
oscillator strength. 

\begin{figure}
\centerline{\psfig{figure=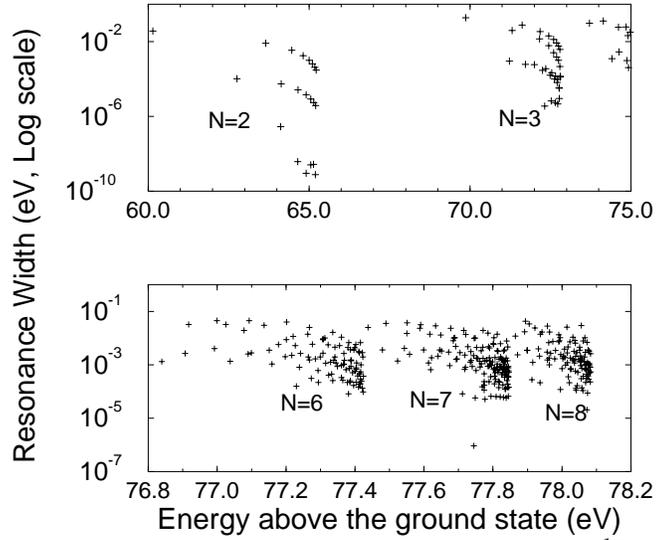,angle=-90,width=10cm}}
\medskip
\caption{\label{pc} Positions (in eV above the ground state) 
and widths (logarithmic scale) of ${}^1\mathbf{P}^o$ resonances of
the Helium atom. The upper plot
displays the states below the $N=2$ and $N=3$ thresholds, where the various series can be
distinguished without ambiguity from their widths, in agreement
with Herrick classification of doubly excited states. In the lower plot, 
displaying the states below the
$N=6,7,8$ thresholds, the various series are strongly coupled and overlapping, which is a quantum manifestation of classical chaos in this system.}
\end{figure}

In the preceding formula, each eigenvalue (resonance
or continuum) contributes to the cross-section at energy $\hbar\omega+E_g$,
with a Fano profile centered at energy $\mathrm{Re}E_i,$ of
width $-2\mathrm{Im}E_i$ whose $q$ parameter is given by~\cite{Buchl94}:
\begin{equation}
q=-\frac{\mathrm{Re}\langle\overline{E_{i\theta}}|R(\theta)T|g\rangle}
{\mathrm{Im}\langle\overline{E_{i\theta}}|R(\theta)T|g\rangle} 
\end{equation}
\noindent Thus, the Fano-$q$ parameter of one resonance is directly 
and unambiguously obtained 
from its associated eigenvector, which is much more efficient
than any fitting procedure, especially above the $N=6$ threshold where 
the different series  strongly overlap (see fig.~\ref{pc}).

For an efficient numerical resolution, the effective Hamiltonian is
expanded in the product of three Sturmian-like basis 
$|n_x\rangle\otimes|n_y\rangle\otimes|n_z\rangle$, 
one for each perimetric
coordinate. The basis states have the following expression:
\begin{equation}
\mathrm{with} \quad \langle u|n\rangle =\phi_n(u)=\sqrt{\alpha_u}
L_n(\alpha_u u)e^{-\alpha_u\frac u2}
\end{equation} 
\noindent where $n_{x,y,z}$ are non negative integers, $\alpha_{x,y,z}$ are real positive
parameters (the scaling
parameters)
and $L_n$ the $n^{\mathrm{th}}$ Laguerre polynomial. This 
non-orthogonal basis is associated
with a representation of the dynamical group SO(2,1), which gives rise
to selection rules. The matrix representation of the effective
Hamiltonian in this basis is thus sparse and banded, and 
the matrix elements are analytically known. 
Let us emphasize that this approach is ``exact"
for the non-relativistic $H\!e$ atom and similar to the one used in reference~
\cite{Wintgen93}.

\begin{figure}
\centerline{\psfig{figure=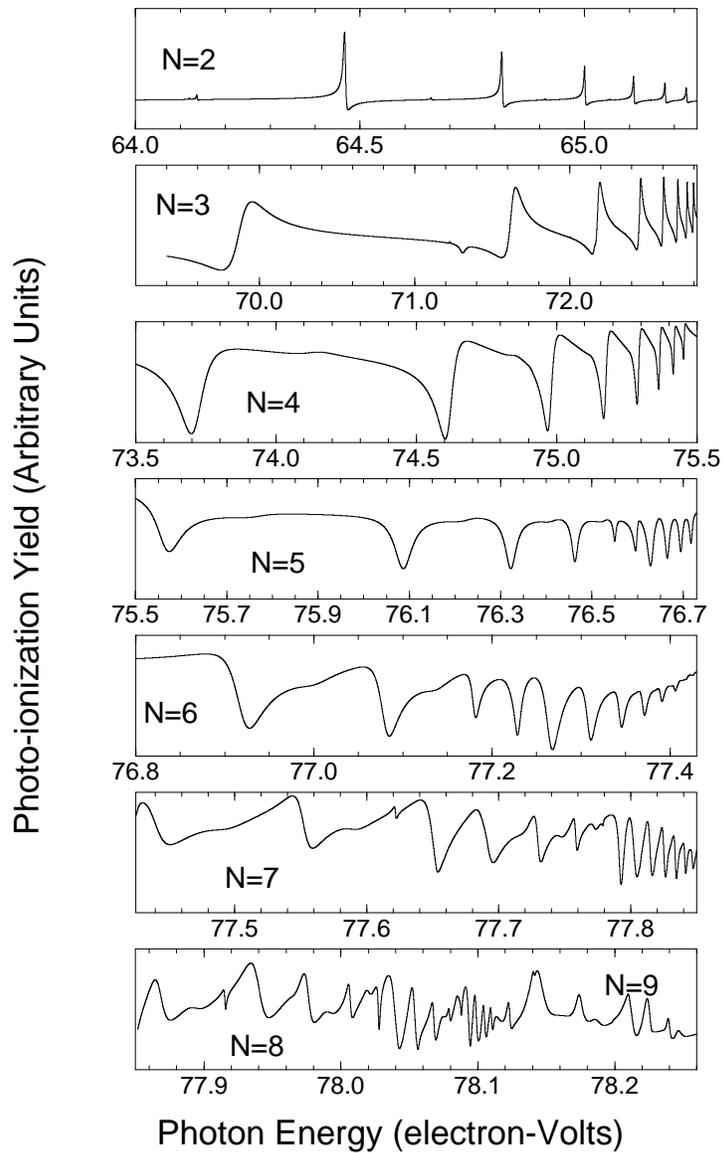,height=15cm}}

\caption{\label{N=2} Calculated photo-ionization cross-section of the helium
atom from the $N=2$ (upper plot) to the $N=8$ and $9$ (lower plot)
series. The raw spectrum has been convoluted by a lorentzian
of width 2 meV for $N=2,3$ and 4 meV for $N=4,5,6$ (equal to the experimental resolution) and $1$ meV for $N=7,8,9$. The calculated cross-section is 
in excellent agreement with the
experimental results of references~\protect\cite{Domke95,Domke96},
displaying for $N=7,8,9$ new peaks, not yet experimentally observed. 
At the highest energies,
the various series overlap strongly, leading to irregular fluctuations
of the cross-section and breakdown of the classification.
Only the fluctuating part of the cross-section is here represented,
the smooth background being subtracted.}
\end{figure}

For obvious reasons, the basis has to be truncated, the prescription
being $n_x+n_y+n_z\leq N_{\mathrm{max}}$ (we used up to $N_{\mathrm{max}}=58$). 
The different scaling parameters are
related by $\alpha=\alpha_x=2\alpha_y=2\alpha_z$, which increases the sparsity
of the matrices and gives the correct decrease for $r_1$ and $r_2$ going to
infinity. The matrices are diagonalized with the Lanczos algorithm,
which is a highly efficient iterative method to obtain few eigenvalues of
huge matrices in a short CPU time~\cite{Ericsson}. 
Convergence of the results are checked
with systematic changes of $\alpha$ and $\theta$. 
We have thus computed few hundred
${}^1\mathbf{P}^o$ states, which are the only ones populated in
one photon experiment starting from the Helium ground state
(${}^1\mathbf{S}^e$). The resulting cross-section  from below the $N=2$ 
up to above the $N=8$ threshold -- the highest energy where experimental spectra
are available -- is shown in figure~\ref{N=2},
convoluted with a lorentzian at the experimental resolution 
($2$ meV for $N=2,3$ and $4$ meV for $N=4,5,6$) 
or at a slightly better resolution (1 meV above the $N=6$ threshold). 
The agreement with the figures from references~\cite{Domke95,Domke96} is 
excellent, emphasizing the efficiency of our calculations. 
The theoretical positions, linewidths and Fano-$q$ parameters are in good
agreement with previous works~\cite{Wintgen93,Domke95}.

\begin{figure}
\centerline{\psfig{figure=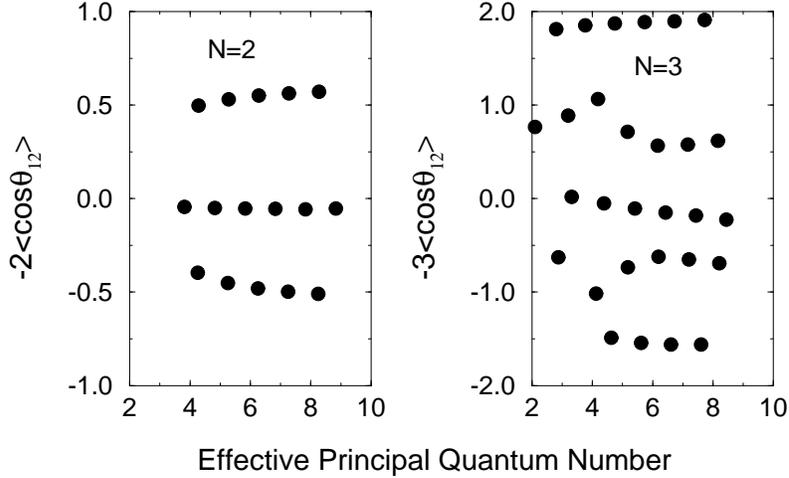,angle=-90,width=10cm}}
\medskip
\caption{\label{cos} Real part of the expectation values of $-N\cos\theta_{12}$ 
for the various resonances below the 
$N=2$ and $N=3$ thresholds. As expected from the 
Herrick classification scheme of doubly excited states, the value
is almost constant across a series, although it
slightly differs from the predicted value $-(N-1)\leq K \leq (N-1).$ 
At higher energy, this classification breaks down.}
\end{figure}

	Below the $N=2$ (resp. $N=3$), three (resp. five) different series are
clearly distinguishable, either by their widths (see fig.~\ref{pc}) or by
the expectation value of $\cos\theta_{12}$, as shown in Fig.~\ref{cos},
where the real part of $-N\cos\theta_{12}$ is plotted 
(the imaginary part is at least ten times smaller)  versus the effective
principal quantum number $n_{\mathrm{eff}}$ of the outer electron
measured from the $N^{th}$ threshold, proving thus the validity of
Herrick classification in these energy ranges.  Still, 
the chaotic aspect of the Helium atom is already observable in the
fluctuations of the smallest widths (see Fig.~\ref{pc}) 
(and also in the Fano-$q$ parameters), 
which will be amplified at higher energies.

\begin{figure}
\centerline{\psfig{figure=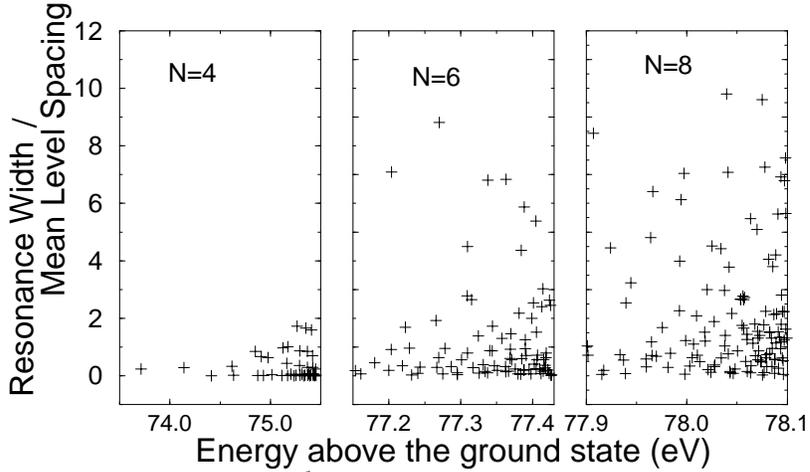,angle=-90,width=10cm}}
\medskip
\caption{\label{ericson} Ratio between the widths $\Gamma$ of
the various ${}^1\mathbf{P}^o$ resonances of the helium atom and the 
local mean level spacing $s$,
displayed for the $N=4$, $N=6$ and $N=8$ thresholds (from left to right). The
transition between the regime of well separated resonances to the strong 
overlapping resonances regime is observed. 
For higher thresholds, the number of resonances
lying above the $\Gamma=s$ line will increase, leading to the observation
of Ericson fluctuations in the photo-ionization cross-section.}
\end{figure}

Below the $N=5$ and $N=6$
thresholds, irregularities due to the interaction with the $6,4_6$ (resp. $7,5_7$) state
from the upper series
are visible, in perfect agreement with the experimental observation.
Below the $N=7$, $N=8$ and $N=9$
thresholds, the computed cross-section -- represented at a better resolution -- 
reproduces very well the various overlapping series, with
an increasing number of perturbers coming from higher series. Furthermore, we show 
new peaks that are not yet experimentally resolved -- such as the members
of the $9,7_n$ series -- but whose observation
could be possible with a (slight) increase of the experimental resolution and
signal to noise ratio. In this energy range, 
the various series are so strongly coupled and overlapping
that the approximate Herrick classification breaks down~\cite{Burgers95,scalhel}, giving rise to an irregular spectrum, showed
by Fig.~\ref{pc} where no general trend can be easily
recognized in the widths of the various resonances. This irregularity is
the quantum manifestation of the chaotic classical dynamics.
In this regime, the photo-ionization cross-section results from the
superposition of various overlapping Fano profiles, eventually leading
to random-like fluctuations in the cross-section known as Ericson fluctuations
\cite{Ericson}. Predicted around the $N=30$ threshold 
in the 1-dimensional helium atom~\cite{Blumel96}, 
this irregular regime takes place at much
lower energy in the real helium atom, because of the increased density of
states. The ratio between the 
linewidth $\Gamma$ and the local mean level spacing $s$ is 
displayed in Fig.~\ref{ericson} for the $N=4$, $N=6$ and $N=8$ thresholds. 
We clearly see
that for $N=8,$ a vast majority of resonances lie above the line 
$\Gamma/s=1$, corresponding to the overlapping resonances regime. 
The published
experimental results \cite{Domke96},
seem to show irregular fluctuations, the first steps towards the Ericson fluctuations.

In conclusion, our results are, as far as we know, the {\em ab initio}
calculations for the double excited ${}^1\mathbf{P}^o$ states of 
the helium atom at the highest energy 
ever done. They are in excellent agreement with the presently
available experimental data. Importantly, they show that the
strongly irregular regime where various resonances overlap
leading to Ericson fluctuations in the photo-ionization cross-section
is almost reached experimentally, which opens the way to their experimental
observation and more generally to a new generation of experiments
probing the
chaotic aspects of the helium atom.

	During this research, B.G. has been financially supported by a
fellowship of the European Commission under contract No. ERBCHBICT941418.
CPU time on a Cray C98 computer has been provided by IDRIS.
Laboratoire Kastler Brossel is laboratoire de l'Universit\'e Pierre et Marie
Curie et de l'Ecole Normale Sup\'erieure, unit\'e associ\'ee 18 du CNRS.

\end{document}